\def \SAIT #1 #2 {{\em Mem.\ Soc.\ Astron.\ It.\/} {\bf #1}, #2}
\def \MESS #1 #2 {{\em The Messenger\/} {\bf #1}, #2}
\def \ASTRNACH #1 #2 {{\em Astron. Nach.\/} {\bf #1}, #2}
\def \AAP #1 #2 {{\em Astron. Astrophys.\/} {\bf #1}, #2}
\def \AAL #1 #2 {{\em Astron. Astrophys. Lett.\/} {\bf #1}, L#2}
\def \AAR #1 #2 {{\em Astron. Astrophys. Rev.\/} {\bf #1}, #2}
\def \AAS #1 #2 {{\em Astron. Astrophys. Suppl. Ser.\/} {\bf #1}, #2}
\def \AJ #1 #2 {{\em Astron. J.\/} {\bf #1}, #2}
\def \ANNREV #1 #2 {{\em Ann. Rev. Astron. Astrophys.\/} {\bf #1}, #2}
\def \APJ #1 #2 {{\em Astrophys. J.\/} {\bf #1}, #2}
\def \APJL #1 #2 {{\em Astrophys. J. Lett.\/} {\bf #1}, L#2}
\def \APJS #1 #2 {{\em Astrophys. J. Suppl.\/} {\bf #1}, #2}
\def \APSS #1 #2 {{\em Astrophys. Space Sci.\/} {\bf #1}, #2}
\def \ASR #1 #2 {{\em Adv. Space Res.\/} {\bf #1}, #2}
\def \BAIC #1 #2 {{\em Bull. Astron. Inst. Czechosl.\/} {\bf #1}, #2}
\def \JSQRT #1 #2 {{\em J. Quant. Spectrosc. Radiat. Transfer\/} {\bf #1}, #2}
\def \MN #1 #2 {{\em Mon. Not. R. Astr. Soc.\/} {\bf #1}, #2}
\def \MEM #1 #2 {{\em Mem. R. Astr. Soc.\/} {\bf #1}, #2}
\def \PLR #1 #2 {{\em Phys. Lett. Rev.\/} {\bf #1}, #2}
\def \PASJ #1 #2 {{\em Publ. Astron. Soc. Japan\/} {\bf #1}, #2}
\def \PASP #1 #2 {{\em Publ. Astr. Soc. Pacific\/} {\bf #1}, #2}
\def \NAT #1 #2 {{\em Nature\/} {\bf #1}, #2}

\documentstyle[twoside,epsfig]{memsait}
\begin{opening}
\title{FIRST RESULTS FROM BeppoSAX}
\author{L. Piro, on behalf of the BeppoSAX team$^1$}
\institute{Istituto Astrofisica Spaziale, C.N.R., Via E. Fermi 21, 00044
Frascati, Italy\\
}
\date{} 
\end{opening}

\begin{document}

\oddpagefooter{}{}{} 
\evenpagefooter{}{}{} 
\ 
\bigskip

\begin{abstract}
The X-ray satellite BeppoSAX
\footnote{ The BeppoSAX team is composed by scientists from:
\begin{itemize}
\item  Istituto Astrofisica Spaziale (IAS), C.N.R., Frascati and Unita'
GIFCO Roma
\item Istituto di Fisica Cosmica ed Applicazioni Informatica (IFCAI), C.N.R.,
 and Unita'
GIFCO, Palermo
\item Istituto Fisica Cosmica e Tecnologie Relative (IFCTR), C.N.R.  and
Unita' GIFCO, Milano
\item Istituto per le Tecnologie e Studio Radiazioni Extraterrestri (ITeSRE),
C.N.R., Bologna and Universita' di Ferrara
\item Space Research Organization of the Netherlands (SRON), The
Netherlands
\item Space Science Department (SSD), ESA, Noordwijk, The
Netherlands
\item BeppoSAX Science Data Center, Rome
\item BeppoSAX Science Operation Center, Rome
\end{itemize}}
, a major programme of the Italian
space agency with participation of the Dutch space agency, was
 successfully launched
from Cape Canaveral on April 30, 1996. After a 2 months  period devoted to
engineering check out that confirmed the nominal functionality of
the satellite and the scientific payload, we have performed a series
of observations of celestial objects devoted to calibrate the instruments
and verify their scientific performances.  Here we will present some
preliminary results obtained in this phase. They are confirming the
expected scientific capabilities of the mission.

\end{abstract}

\section{Introduction}
The X-ray satellite SAX, named BeppoSAX after launch in honour
of Giuseppe (Beppo) Occhialini, is the first X-ray mission
with a scientific payload covering more
than three decades of energy - from 0.1 to 300 keV - with a relatively
large area, a good energy resolution, and with imaging capabilities
(resolution of about 1 arcmin) in the range of 0.1-10 keV.  This
capability, in conjunction with the presence of wide field instruments
primarily aimed at discovering transient phenomena, which could then be
observed with the broad band instruments, provides an unprecedented
opportunity to study the broad band behaviour of several classes of
X-ray sources.

The broad band capability is provided by a set of
instruments co-aligned with the Z axis of the satellite,
Narrow Field Instruments (hereafter NFI) and
composed by:
\begin{itemize}
\item MECS (Medium Energy Concentrator Spectrometers):
a medium energy (1.3-10 keV)
 set of three identical grazing incidence telescopes with double cone geometry
(Citterio et al. 1985, Conti et al. 1994),
 with position sensitive gas scintillation proportional counters
in their focal planes (\cite{mecs}).
\item LECS (Low Energy Concentrator Spectrometer):
a low energy (0.1-10 keV) telescope,
identical to the other three, but with a thin window position
sensitive gas scintillation proportional counter in its focal plane
(\cite{lecs}).
\item HPGSPC, a collimated High Pressure Gas Scintillation
 Proportional Counter (4-120 keV, \cite{hpgspc}).
\item PDS, a collimated Phoswich Detector System (15-300 keV,
\cite{pds})
\end{itemize}

Access to large regions of the sky ($\sim3000$ degree$^2$) with a
resolution of 5' in the range 2-30 keV is provided by:
\begin{itemize}
\item two coded mask proportional counters (Wide Field Cameras, WFC,
\cite{wfc}), perpendicular to the axis of the NFI and pointed in
opposite directions.

\end{itemize}

Finally, the anticoincidence scintillator shields of the PDS (GRBM) will be
used as a gamma-ray burst monitor in the range 60-600 keV with a
fluence greater than about $10^{-6}$ erg cm$^{-2}$ and with a temporal
resolution of about 1 ms.

More details on the mission and its instruments can be found in
\cite{psb95}, \cite{general}, in the special session devoted to
BeppoSAX of the SPIE Vol. 2517 and on line at:

{\it http://www.sdc.asi.it.}

\section{THE SCIENCE VERIFICATION PHASE}
The goal of the Science Verification Phase is to verify the expected
scientific capabilities of the mission and to calibrate the
instruments.  To this aim a series of objects with well known
properties have been selected and are being observed. Here we present
some preliminary results on the X-ray pulsar Vela X-1, the Seyfert
galaxy NGC 4151 as examples of the broad band spectroscopy of bright
and weak sources,  and
the simultaneous observation with the WFC and the GRBM of a
$\gamma$-ray burst to show the capabilities of monitor wide regions of
the sky to  observe transient phenomena.
\begin{figure}
  \begin{center} \leavevmode
\psfig{file=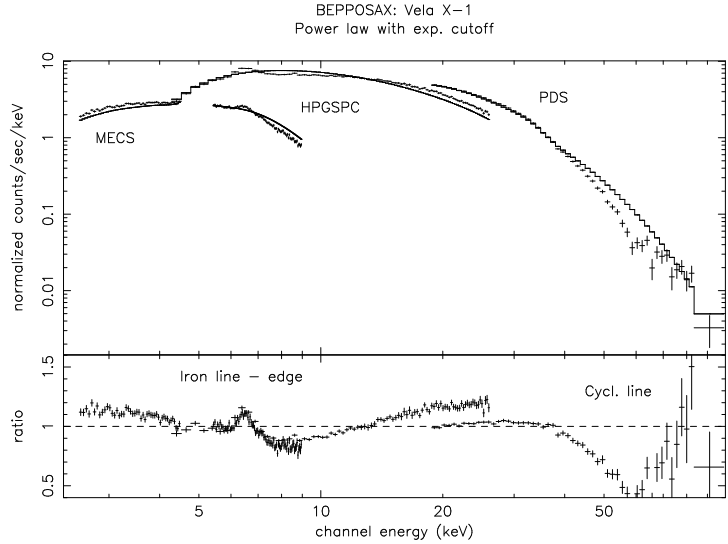,
    angle=270, width=11cm} \end{center} \caption{ The pulse height spectrum of
    the X-ray pulsar Vela~X-1 by the MECS, HPGSPC and PDS fitted with
    a power law with exponential cut-off. The residuals show the
    presence of iron line and absorption edge as well as an absorption
    feature around 60 keV} 
\end{figure}
\begin{figure}
  \begin{center} \leavevmode
\epsfig{file=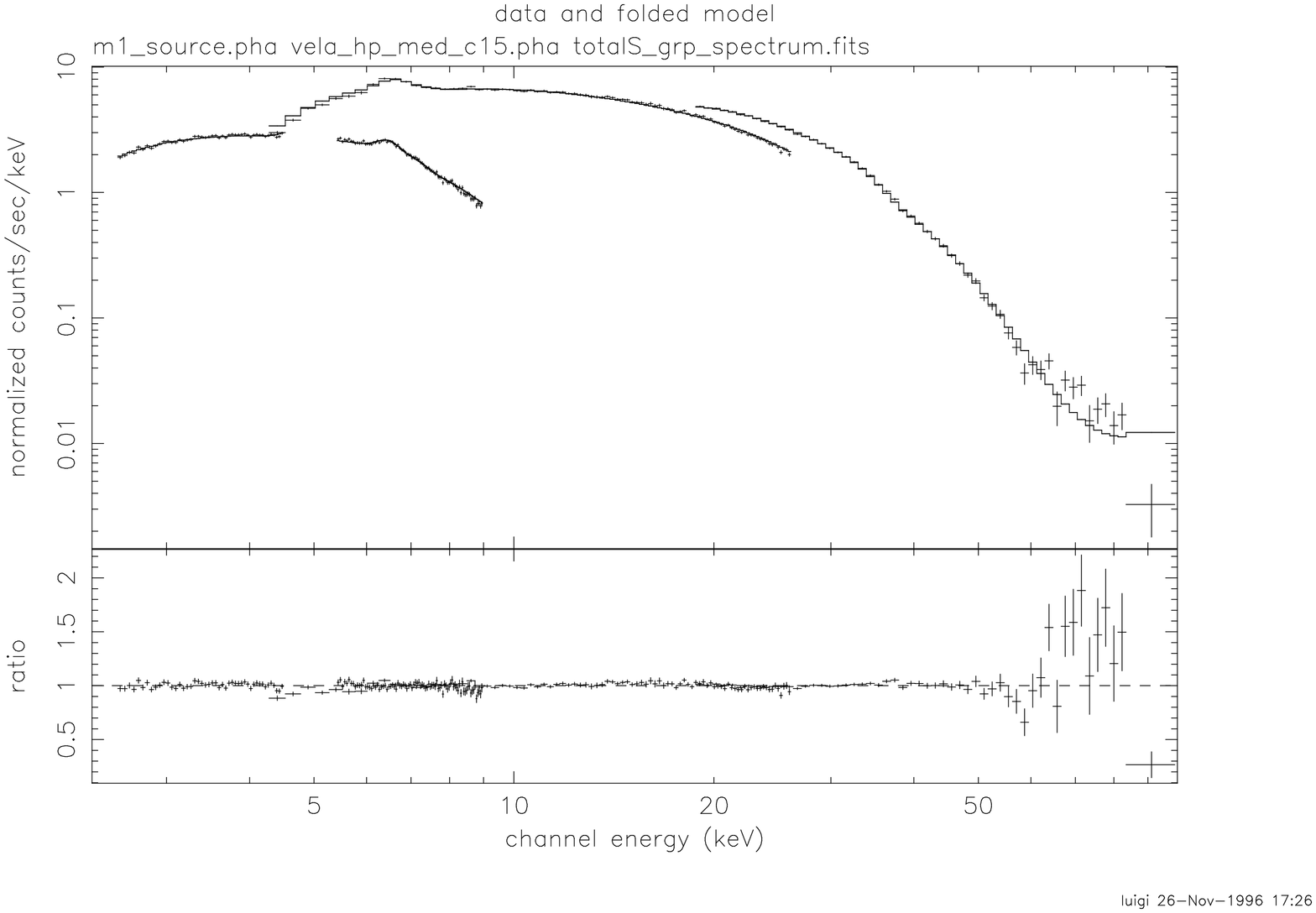,
    angle=270, width=11.0cm,
 bbllx=20pt , bblly=20pt, bburx=560pt, bbury=774pt,
    clip=  } \end{center} \caption{The best fit to the
    Vela X-1 spectrum obtained with a power law with 2 cyclotron lines
    at about 30 and 60 keV, iron line and absorption edge (see text) }
\end{figure}
\subsection{ Broad Band Spectroscopy with the NFI}
\subsubsection{The X-ray pulsar Vela X-1}
In fig.1 we show the spectrum of the X-ray pulsar Vela X-1 in the
range 3-200 keV obtained in a 30 ksec observation with the MECS,
HPGSPC and PDS.  The data are fitted with a power law with absorption
and an exponential cut off.  The residuals show large deviations from
the model. The most noticeable are those due to the presence of an
iron line and absorption edge in the 6-8 keV region and an absorption
feature around 60 kev.  Following the results from GINGA (Mihara 1995)
we have then fitted the spectrum with a power law with two cyclotron
absorption lines, plus an iron line and iron edge. This model provides
a satisfactory fit to the data (fig.2).  The values of the cyclotron
lines are remarkably similar to those obtained by Mihara in a fit employing the
same model. The first line is at around 27 keV, with an optical
depth of about 0.2, whereas the optical depth of the second harmonic
at 54 keV is about 10 times larges. The two lines are rather broad,
being respectively about 15 keV and 35 keV. Further analysis is on going
to study different models and phase-resolved spectra.

\begin{figure}
  \begin{center} \leavevmode
 \epsfig{file=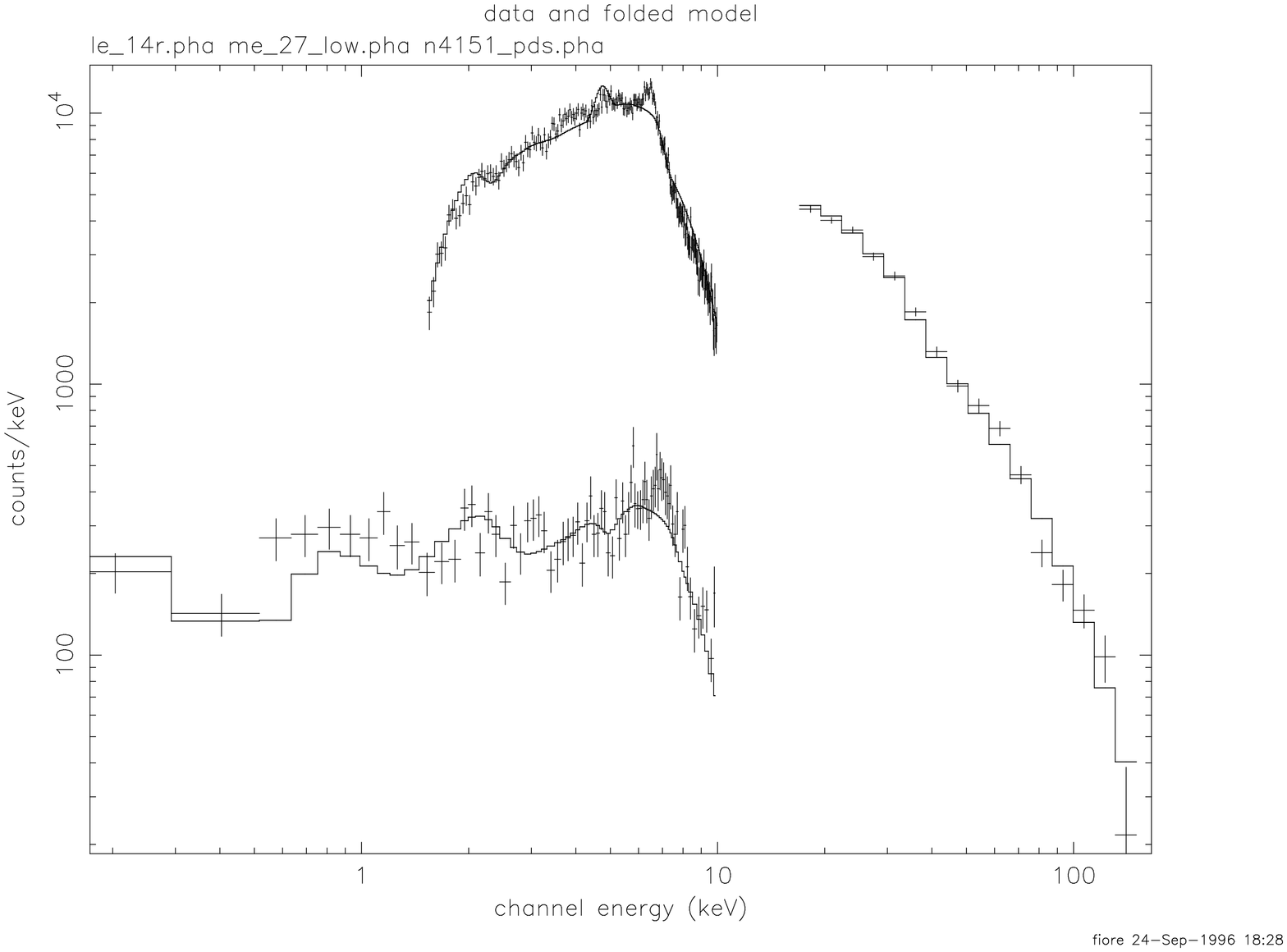,
    width=8cm , bbllx=22pt , bblly=20pt, bburx=570pt, bbury=774pt,
    clip= , angle=270} \end{center} \caption{ The pulse height spectrum (detector cts /s) of
    NGC~4151 observed by the LECS, MECS and PDS (from left to right)
    fitted with a complex model: a power law, a soft X-ray component,
    a complex absorbing medium, a high energy cut-off}
\end{figure}

\begin{figure}
  \begin{center} \leavevmode
\epsfig{file=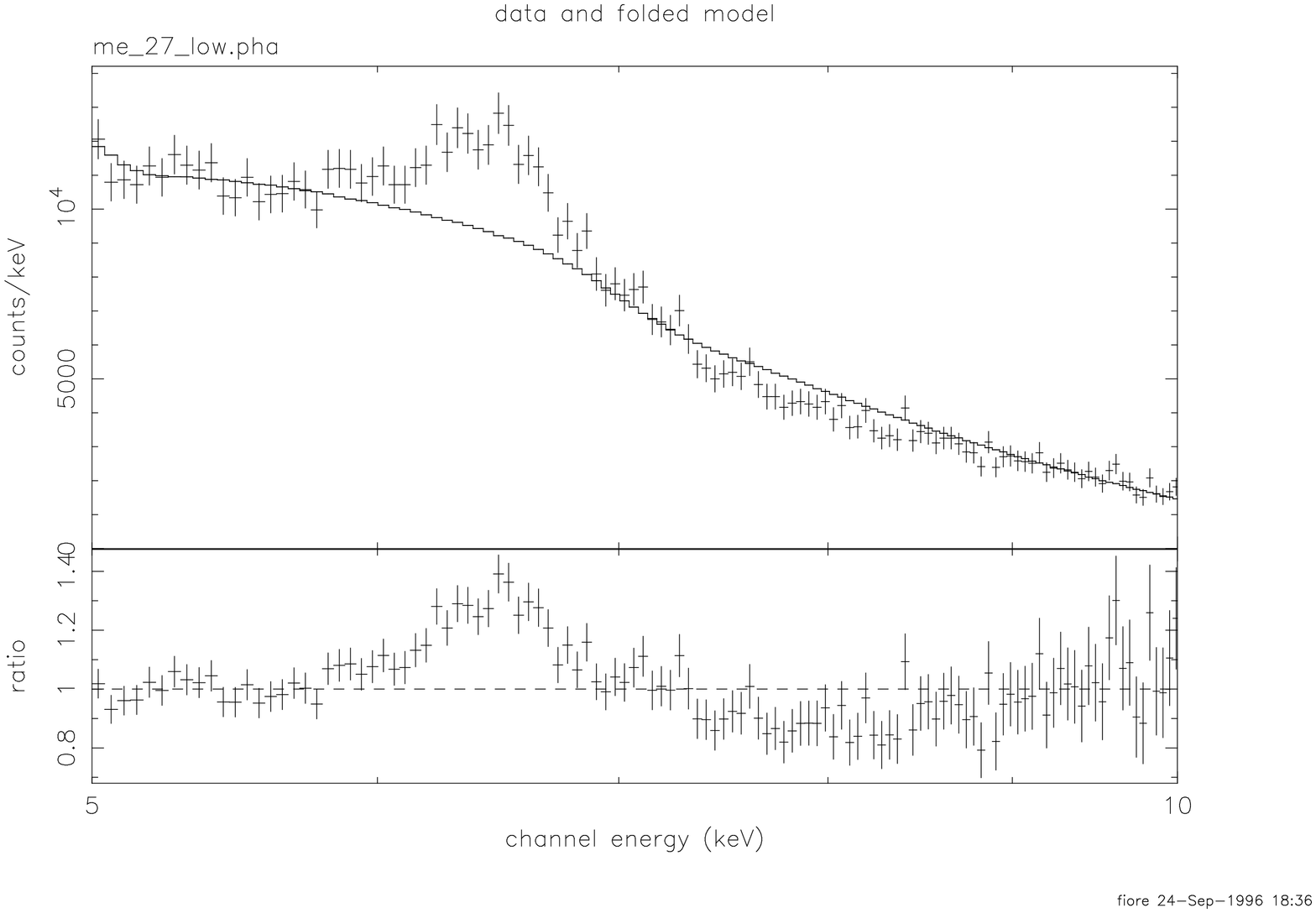, 
    width=6cm , bbllx=20pt , bblly=20pt, bburx=560pt, bbury=774pt,
    clip= , angle=270} \end{center} \caption{ The spectrum of NGC
    4151 in the MECS around the iron complex region. The best fit
    model is the same as that in fig.3 to show in the residuals the
    clear presence of the iron line and absorption edge}
\end{figure}
\begin{figure}
  \begin{center} \leavevmode
\epsfig{file=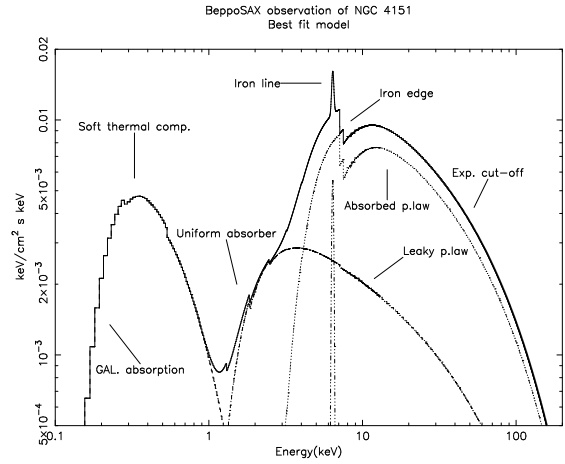, 
    width=6cm , bbllx=20pt , bblly=20pt, bburx=560pt, bbury=774pt,
 angle=270} \end{center} \caption{ 
Best fit model (photon spectrum) of NGC 4151} 
\end{figure}

\begin{figure}
  \begin{center} \leavevmode
\epsfig{file=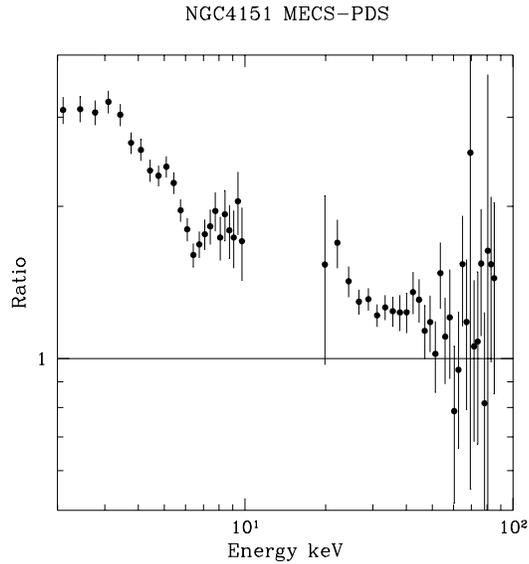, 
width=8cm, bbllx=20pt, bblly=180pt, bburx=560pt, bbury=774pt, trim= -10 0 -30 0, 
clip= 
} \end{center} 
\caption{ Ratio of the high and low state MECS and PDS spectra of NGC
4151. The spectral variability extends above 10 keV, and is therefore
most likely produced by a change in the slope of the intrinsic
continuum. Note as the region around 6.4 keV remains relatively
constant, indicating that the iron line does not follow continuum
variations on a scale of a few days.}

\end{figure}


\begin{figure}
  \begin{center}
\epsfig{file=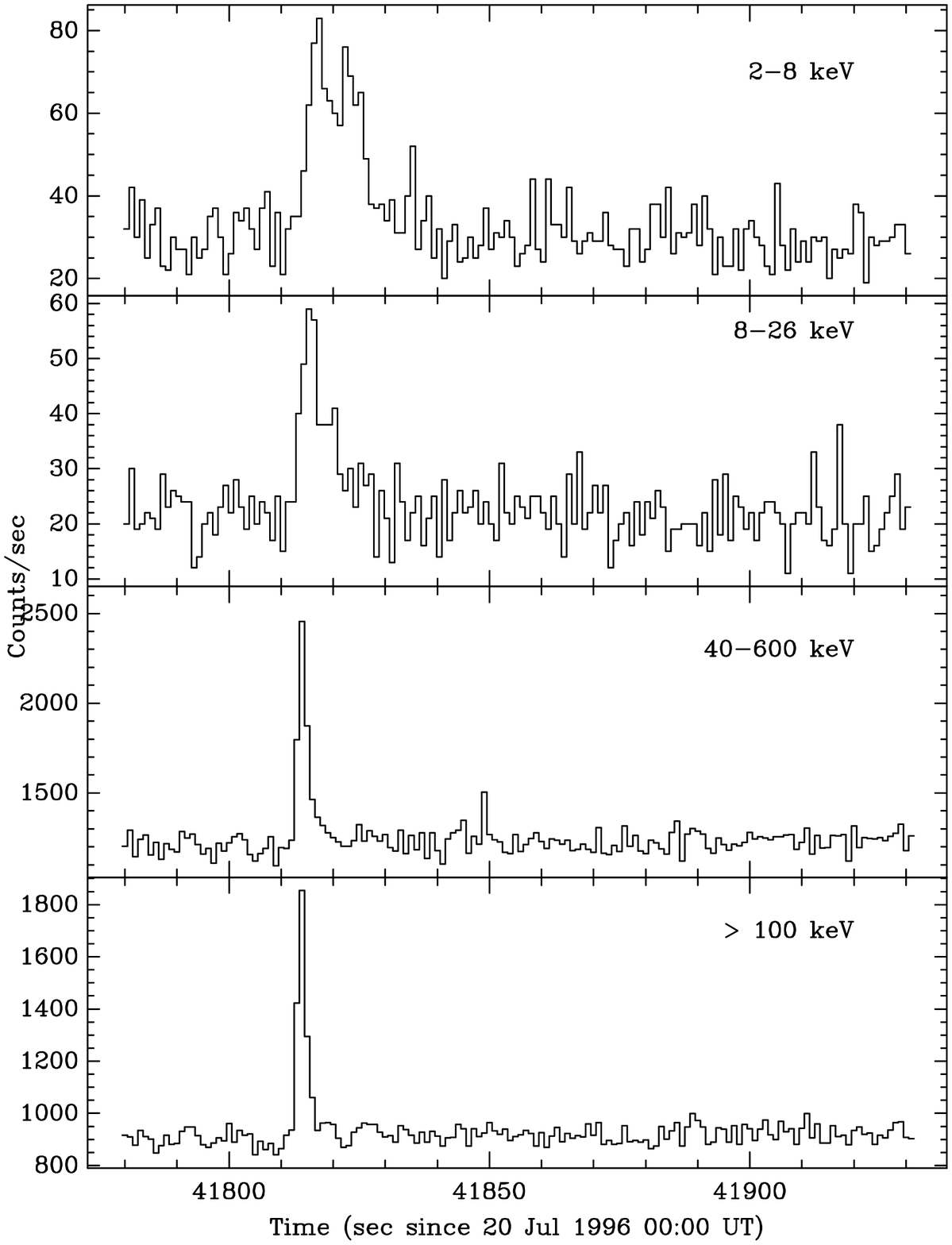,
 width=7.0cm,
 bbllx=20pt
, bblly=20pt,
  bburx=560pt, bbury=774pt, clip=}
  \end{center}
  \caption{ Light curves of the $\gamma$-ray burst (GB960720) detected
simultaneously in the GRBM and WFC detectors}
  \label{fig:echbasegaia}
\end{figure}

\subsubsection{The Seyfert 1 galaxy NGC 4151: the spectrum}

The X-ray spectrum of NGC 4151 is the most complex observed so far in
AGN, being characterized by narrow and broad spectral features from
soft to hard X-rays (e.g. \cite{petal86}; \cite{wds95}; \cite{zjm96}).
It is then the
best candidate to verify the unique capability of BeppoSAX to
disentangle spectral features over the 0.1-200 keV energy range.

In fig.3 we show the spectrum of the LECS (7 ksec of effective
exposure time), MECS (about 55 ksec) and PDS (about 35 ksec) fitted
with a complex model of the broad continuum components. The presence
of iron line and iron absorption edge is very clear in the residuals
of the MECS (fig.4).  In fig.5 we show the best fit model spectrum
required to fit the data and composed by : an intrinsic power law with
an exponential cut-off around 70 keV; an absorbing medium with a
column density $\sim 10^{23} cm^{-2}$ which is likely producing the
observed iron fluorescence line and the iron absorption edge; this
medium has a complex structure, well described by a leaky absorber,
that allows a fraction ($\sim 20\%$) of the intrinsic power law
continuum to be transmitted without strong absorption.  However, to be
consistent with the spectrum (and lack of variability, see
\cite{petal86}) in soft X-rays, this component needs to be absorbed by
a further, uniform absorbing screen with $N_H\sim 10^{22} cm^{-2}$.
Finally, a soft component, possibly of thermal origin ($kT\sim $0.4
keV), external to the uniform absorber, is present below 1 keV.

\subsubsection{The Seyfert 1 galaxy NGC 4151: spectral variability}
The study of spectral variability provides the most powerful tool to
identify the origin of the several components - some of which
presumably share the same physical origin - located across the whole
X-ray band. In the past this approach has been hampered basically by
the limited bandwidth of the instruments, that did not allow to
disentangle the different contribution of those components to the
observed spectral variability.

From this point of view BeppoSAX provides an unique opportunity and
the case of NGC 4151 is a very good example of its capability. EXOSAT
observations of this object showed a clear evidence of spectral
variability in the band 2-10 keV. However its origin was attributed
either to a change in the slope of the intrinsic continuum (Perola et
al. 1986) or to a variation associated with the absorbing medium
(Yaqoob et al. 1989). GINGA observations suggested that, of the two
explanations, an intrinsic variation was preferred by the data (Yaqoob
and Warwick 1991).

Here we show some of the results we obtained during the observation by
BeppoSAX. The object increased its flux by a factor of 2-3 in about 2
days. In fig.~6 we show the ratio of the spectra of the MECS and PDS
relative to the second and the first part of the observation, when the
source was respectively in a high and low state.  The spectral change
is evident in the whole range. In particular we note that the spectrum
changes its shape also above 20 keV, a region where any effect of
absorption by a medium with column densities as observed in this
object ($N_H\sim 10^{23}\ $ cm$^{-2}$) is negligible.

 Another result that appears from fig.~6 is the smaller variation
observed around 6.4 keV (see also Molendi et al. 1997) indicating that
the iron line does not follow the continuum variations on a scale of a
few days. The medium producing the iron line has thence to be located
at greater distances from the central source (Perola et al. 1986).

\subsection{Transient phenomena: the WFC and the Gamma-ray burst monitor}
One of the primary scientific goals of BeppoSAX is the observation of
transient phenomena in the sky. Two set of instruments are devoted to
this purpose: the two WFC and the GRBM. 

An exciting example of the capability of BeppoSAX of observing
 transient phenomena is the simultaneous observation of
the $\gamma$-ray burst GB960720 by the WFC and the GRBM
(Piro et al. 1996a). The WFC image allows to localize the event
within a few arcmin. This observation has triggered a series of follow up
observations in different energy bands (e.g. Frail et al. 1996; 
Murakami et al. 1996). We have carried out a deep
observation of the field with the NFI, that has led to the
discovery of a previously unknown X-ray source in the WFC error box
(Piro et al. 1996b). It is not yet clear whether this source is
actually related with GB960720, since the probability of finding a serendipitous AGN at this flux level in the error box is of about 0.1. Along with the imaging information,
the wide band energy range covered simultaneously by the WFC and the
GRBM provides important information of the evolution of the burst
at different energies.
In fig.~7 we show the light curves of the event in different energy
ranges of the two instruments. The event duration decreases at higher energies.
On the basis of the logN-logS of the $\gamma$-ray bursts and the
instruments' sensitivities we expect to detect about 5 $\gamma$-ray burst
 in a year in both instruments.


\end{document}